\pdfoutput=1
\documentclass[pra,twocolumn,showpacs,superscriptaddress,hyperref]{revtex4}%
\usepackage{amsfonts}
\usepackage{hyperref}
\usepackage{amsmath}
\usepackage{amssymb}
\usepackage{graphicx}%
\providecommand{\U}[1]{\protect\rule{.1in}{.1in}}

\begin{document}
\title{Sudden change in quantum and classical correlations and the Unruh effect}
\author{L. C. C\'{e}leri}
\email{lucas.celeri@ufabc.edu.br}
\affiliation{Centro de Ci\^{e}ncias Naturais e Humanas, Universidade Federal do ABC, Rua
Santa Ad\'{e}lia 166, 09210-170, Santo Andr\'{e}, S\~{a}o Paulo, Brazil}
\author{A. G. S. Landulfo}
\email{landulfo@ift.unesp.br}
\affiliation{Instituto de F\'{\i}sica Te\'{o}rica, Universidade Estadual Paulista, Rua Dr.
Bento Teobaldo Ferraz, 271-Bl. II, 01140-070 S\~{a}o Paulo, S\~{a}o Paulo, Brazil}
\author{R. M. Serra}
\email{serra@ufabc.edu.br}
\affiliation{Centro de Ci\^{e}ncias Naturais e Humanas, Universidade Federal do ABC, Rua
Santa Ad\'{e}lia 166, 09210-170, Santo Andr\'{e}, S\~{a}o Paulo, Brazil}
\author{G. E. A. Matsas}
\email{matsas@ift.unesp.br}
\affiliation{Instituto de F\'{\i}sica Te\'{o}rica, Universidade Estadual Paulista, Rua Dr.
Bento Teobaldo Ferraz, 271-Bl. II, 01140-070 S\~{a}o Paulo, S\~{a}o Paulo, Brazil}

\begin{abstract}
We use the Unruh effect to analyze the dynamics of classical and quantum
correlations for a two-qubit system when one of them is uniformly accelerated
for a finite amount of proper time. We show that the quantum correlation is
completely destroyed in the limit of infinite acceleration, while the
classical one remains nonzero. In particular, we show that such correlations
exhibit the so-called \emph{sudden-change} behavior as a function of
acceleration. Eventually, we discuss how our results can be interpreted when
the system lies in the vicinity of the event horizon of a Schwarzschild black hole.

\end{abstract}

\pacs{03.65.Ud, 03.30.+p, 03.67.-a, 04.62.+v}
\maketitle

\section{Introduction}

\label{Introduction}

What is \emph{quantum} in a quantum correlated system? Until recently, the
usual answer to this question was \emph{entanglement}. However, as pointed out
by Ollivier and Zurek~\cite{OllZur} (see also Ref.~\cite{HenVed}), separable
states may still present quantum correlations. Recent results showing that
almost all (separable) quantum states actually have such nonclassical
correlations~\cite{Ferraro} and that they can improve some computational tasks
in comparison to when classical states are used~\cite{Caves,White} reinforce
the relevance of such an issue. As a result, the last few years have witnessed
an increasing number of articles discussing the quantification of these
correlations~\cite{OllZur,HenVed, DiVincenzo,Lou1,Opp}, their behavior under
decoherence~\cite{Maziero,Maziero2,Werlang,Fanchini}, and their relevance for
quantum phase transitions~\cite{Dillenschneider,Sarandy,Werlang2,Maziero3}.

Parallel to these developments, theoretical studies analyzing the role played
by relativity in the behavior of quantum systems when \textbf{(i)}~the
relative motion and \textbf{(ii)}~proper acceleration of the communicating
partners is large have attracted much attention (see Ref.~\cite{Peres} for a
recent review). This is not only interesting in its own right but may also
have practical importance because of new trends of implementing
quantum-information protocols at global scales through the use of satellite
systems~\cite{Zeilinger,Pan,Weinfurter,Zeilinger2, Zeilinger3}. Concerning
condition~\textbf{(i)}, the remarkable fact was shown in Ref.~\cite{STR} that
the von Neumann entropy associated with the reduced spin-density matrix of a
single particle is not Lorentz invariant. This is so because, in general, the
\emph{spin} is Wigner rotated under Lorentz boosts in a direction which
depends on the particle \emph{momentum}, thus, entangling both degrees of
freedom. Similarly, it was shown that the entanglement for a two-particle spin
system is not Lorentz invariant either~\cite{STRI}. The Lorentz invariance of
the entanglement distillability of a bipartite mixed spin state was
investigated in Ref.~\cite{Lamata}. The degree of violation of the
Clauser-Horne-Shimony-Holt inequality as seen by different inertial observers
for two entangled fermions was studied in Ref.~\cite{LM09} and for entangled
photons in Ref.~\cite{LMT10}. Concerning condition~\textbf{(ii)}, it was shown
in Ref.~\cite{Mann} that uniformly accelerated observers see a degradation in
the degree of entanglement between two bosonic modes in comparison to inertial
observers. This fact can be explained through the well-known Unruh
effect~\cite{Unruh}, which states that uniformly accelerated observers in
Minkowski space-time associate a thermal bath of Rindler particles to the
Minkowski vacuum (see, e.g., Ref.~\cite{MatsasRev} for a review). The case of
continuous variables was studied in Ref.~\cite{Adesso}, where it was shown
that the entanglement is completely destroyed in the limit of infinite
acceleration for scalar fields, in contrast to what was found for Dirac
ones~\cite{Tessier}. In a recent work, two of the present authors observed
sudden death of entanglement in a two-qubit system when one of the partners is
accelerated by some external agent~\cite{LM209}. A similar result for
circularly moving qubits can be found in Ref.~\cite{DC10}.

A common feature shared by the previous articles is that all of them consider
some entanglement measure to quantify the quantum correlation contained in
composite systems. Here we are interested in the dynamics of a more general
class of (classical and quantum) correlations as given, for example, by the
\emph{quantum discord}~\cite{OllZur}. For this purpose we consider a two-qubit
system where one of the partners is uniformly accelerated for a finite amount
of proper time. We verify that the quantum and classical correlations exhibit
a \emph{sudden-change} behavior~\cite{Maziero} as a function of the
noninertial qubit acceleration. In addition, we show that the quantum
correlation is completely destroyed in the limit of infinite acceleration in
contrast to the classical one. An analysis that complement ours can be found
in Ref.~\cite{Datta}, where the author studies the quantum discord between two
\emph{free} modes from the point of view of relatively accelerated observers.
He finds that the quantum correlation between the modes is not entirely
destroyed from the point of view of these observers even in the limit of
infinite acceleration. It is important to note that the setup in
Ref.~\cite{Datta} differs from ours, where one of the qubits is uniformly
accelerated and thus under the influence of some external force.

The article is organized as follows. In Sec.~\ref{System}, we introduce our
two-qubit system. In Sec.~\ref{Correlations}, we discuss some quantifiers of
classical and quantum correlations. Section~\ref{Correlation dynamics} is
dedicated to presenting our results on the behavior of the correlations, which
are also discussed in terms of the Unruh effect. In Sec.~\ref{Black hole}, we
explain how our results can be straightforwardly applied when the qubits are
in the vicinity of the event horizon of a black hole. Our final remarks are
presented in Sec.~\ref{Final}. Throughout this article all logarithms are
taken to base $2$ and we adopt natural units: $\hbar=c=G=k_{B}=1$.


\section{Alice-Rob problem}

\label{System}

In this section, we briefly introduce the qubit system considered here. (We
refer to Ref.~\cite{LM209} for more details.)

\subsection{Qubit Model }

Our qubits are modeled as two-level semiclassical detectors. The semiclassical
character of the detector, usually named after Unruh-DeWitt, relies on the
fact that, while it possesses a well-defined \emph{classical} world line, its
internal degrees of freedom are treated \emph{quantum mechanically}
\cite{MatsasRev}. The detector proper Hamiltonian is defined as
\begin{equation}
H_{D}=\Omega D^{\dag}D,\label{HD}%
\end{equation}
where $\Omega>0$ is the detector energy gap and $D^{\dag}$, $D$ are the
transition operators for the qubit energy eigenstates: $D^{\dag}\left\vert
1\right\rangle =D\left\vert 0\right\rangle =0$, $D^{\dag}\left\vert
0\right\rangle =\left\vert 1\right\rangle $, and $D\left\vert 1\right\rangle
=\left\vert 0\right\rangle $. Here $\left\vert 0\right\rangle $ and
$\left\vert 1\right\rangle $ represent the unexcited and excited qubit energy
eigenstates, respectively.

We coupled the qubit to a massless scalar field operator $\phi(x)$ through the
interaction Hamiltonian
\begin{equation}
H_{I}=\epsilon(t){\displaystyle\int\limits_{\Sigma_{t}}}d^{3}\mathbf{x}%
\sqrt{-g}\phi(x)\left[  \psi(\mathbf{x})D+\psi^{\ast}(\mathbf{x})D^{\dag
}\right]  ,\label{HI}%
\end{equation}
where $g=\det(g_{ab})$, $g_{ab}$ is the Minkowski space-time metric, and
$\mathbf{x}$ are coordinates defined on the Cauchy surface $\Sigma_{t=const}$
associated with the suitable timelike isometries followed by the qubits
(namely, the inertial and uniformly accelerated ones). The smooth compact
support real-valued function $\epsilon$ is introduced to keep the detector
switched on for a finite amount of proper time $\Delta$, which will be always
assumed to have a fixed nonzero value here, and $\psi$ is a smooth compact
support complex-valued function modeling the fact that the detector only
interacts with the field in a neighborhood of its world line. The total
Hamiltonian can be cast as
\begin{equation}
H=H_{0}+H_{I},\label{HT}%
\end{equation}
where $H_{0}=H_{D}+H_{KG}$ and $H_{KG}$ is the free scalar field Hamiltonian.
By using the interaction picture, the final state of the qubit-field system in
first perturbation order is given by~\cite{Wald}
\begin{equation}
|\Psi_{\infty}\rangle=(I+a^{\dagger}(\lambda)D-a(\overline{\lambda}%
)D^{\dagger})|\Psi_{-\infty}\rangle,\label{primeira_ordem_2}%
\end{equation}
where $|\Psi_{-\infty}\rangle$ is the corresponding initial state,
$\lambda=-KEf$, $f=\epsilon(t)\psi(\mathbf{x})e^{-i\Omega t}$, $E$ is the
difference between the advanced and retarded Green functions, $K$ is an
operator that takes the positive frequency part of the solutions of the
Klein-Gordon equation with respect to the timelike isometry, and
$a(\overline{u})$ and $a^{\dagger}(u)$ are the annihilation and creation
operators of $u$ modes, respectively. Equation~(\ref{primeira_ordem_2})
carries the physical message that the excitation and de-excitation of an
Unruh-DeWitt detector following a timelike isometry is associated with the
absorption and emission, respectively, of a particle as \textquotedblleft
naturally\textquotedblright\ defined by observers co-moving with the detector.
Only processes where the detector flips once or none at all are considered here.

\subsection{Alice, Rob and the Unruh effect }

Let us now consider two non-interacting qubits in Minkowski space-time. The
one carried by Alice is kept inertial while the one carried by Rob has
constant proper acceleration $a$ along the $x$ axis for the finite amount of
proper time $\Delta$. The world line of Rob's qubit is given by
\[
t(\tau)=a^{-1}\sinh a\tau,\;x(\tau)=a^{-1}\cosh a\tau,
\]
$y(\tau)=z(\tau)=0$, where $\tau$ is the qubit proper time and $(t,x,y,z)$ are
the usual Cartesian coordinates of Minkowski space-time. This is meaningful as
long as we consider space-localized qubits, which can be realized by choosing
$\psi(\mathbf{x})=(\kappa\sqrt{2\pi})^{-3}\exp(-\mathbf{x}^{2}/2\kappa^{2})$
with variance $\kappa=\mathrm{const}\ll\Omega^{-1}$.

The initial state describing the complete system composed of the two qubits
and scalar field is denoted as
\begin{align}
|\Psi_{0}^{AR\phi}\rangle &  =|\Psi_{0}^{AR}\rangle\otimes|0_{M}%
\rangle\nonumber\\
&  =\frac{1}{\sqrt{2}}(|0_{A}\rangle\otimes|1_{R}\rangle-|1_{A}\rangle
\otimes|0_{R}\rangle)\otimes|0_{M}\rangle\label{IS}%
\end{align}
where $\{|0_{X}\rangle,|1_{X}\rangle\}$ is an orthonormal basis of the
internal qubit space; $X=A$ and $R$ labels the Alice and Rob qubits,
respectively; and $|0_{M}\rangle$ is the scalar field Minkowski vacuum, that
is, the no-particle state as defined by inertial observers.

The free Hamiltonian for each qubit is given by Eq.~(\ref{HD}) with the
appropriate substitution $D\rightarrow A,R$. They are designed such that
Alice's qubit remains always switched off while Rob's qubit is kept switched
on during the nonzero time interval $\Delta$ along which it interacts with the
field through the effective coupling constant $\nu^{2}$ [see Eq.~(\ref{nu})].
As a result, Rob's qubit interacts with the scalar field as ruled by
Eq.~(\ref{HI}) (with $D\rightarrow R$ and $t\rightarrow\tau$), while Alice's
qubit interacts with the scalar field only indirectly. Then, the total
Hamiltonian of the complete two-qubit system interacting with the field is
given by
\begin{equation}
H=H_{A}+H_{R}+H_{KG}+H_{I}. \label{HF}%
\end{equation}
Now, by using Eq.~(\ref{primeira_ordem_2}) to evolve our initial
state~(\ref{IS}) and tracing out the field degrees of freedom eventually, we
obtain the final reduced density matrix~\cite{LM209}
\begin{equation}
\rho_{\infty}^{AR}=\left[
\begin{array}
[c]{cccc}%
S_{2} & 0 & 0 & 0\\
0 & S_{0} & -S_{0} & 0\\
0 & -S_{0} & S_{0} & 0\\
0 & 0 & 0 & S_{1}%
\end{array}
\right]  \label{FS}%
\end{equation}
associated with the two-qubit degrees of freedom in the basis
\[
\{|0_{A}\rangle\otimes|0_{R}\rangle,|1_{A}\rangle\otimes|0_{R}\rangle
,|0_{A}\rangle\otimes|1_{R}\rangle,|1_{A}\rangle\otimes|1_{R}\rangle\},
\]
where
\begin{align*}
S_{0}  &  =\frac{1-q}{2(1-q)+\nu^{2}(1+q)},\\
S_{1}  &  =\frac{\nu^{2}q}{2(1-q)+\nu^{2}(1+q)},\\
S_{2}  &  =\frac{\nu^{2}}{2(1-q)+\nu^{2}(1+q)}.
\end{align*}
For the sake of convenience, we have defined the parametrized acceleration
$q\equiv e^{-2\pi\Omega/a}$, as well as the effective coupling
\begin{equation}
\nu^{2}\equiv||\lambda||^{2}=\frac{\epsilon^{2}\Omega\Delta}{2\pi}%
e^{-\Omega^{2}\kappa^{2}}. \label{nu}%
\end{equation}
We note that a couple of necessary conditions for the relations above to be
valid is that $\Omega^{-1}\ll\Delta$ and that $\epsilon$ be a slow varying
function of time when compared to the frequency $\Omega$.

For the asymptotic limit of infinite acceleration ($q\rightarrow1$), the final
reduced density matrix of the Alice-Rob system turns out to be [see
Eq.~(\ref{FS})]%
\begin{equation}
\left.  \rho_{\infty}^{AR}\right\vert _{q\rightarrow1}=\frac{1}{2}|0_{A}%
0_{R}\rangle\langle0_{A}0_{R}|+\frac{1}{2}|1_{A}1_{R}\rangle\langle1_{A}%
1_{R}|. \label{assimpt}%
\end{equation}
The asymptotic state~(\ref{assimpt}) is a consequence of the assumption that
the detector is allowed to flip only once or none at all~\cite{LM209} and the
fact that in the infinite acceleration limit Rob's detector must necessarily
flip; no flip is not an option in this case. Because each detector excitation
and de-excitation in the usual inertial vacuum is necessarily associated with
the emission of a Minkowski particle, inertial observers must discard data
coming from experiments where two or more Minkowski particles are eventually
left by using some post-selection process. Accordingly, uniformly accelerated
observers must also discard data associated with two or more detector transitions.


\section{Quantum and Classical Correlations}

\label{Correlations}

In classical information theory, the correlation between two random variables
$A$ and $B$ is measured by the mutual information~\cite{CIT}
\begin{equation}
I_{c}(A:B)=\mathcal{H}(A)+\mathcal{H}(B)-\mathcal{H}(A,B),\label{CMI}%
\end{equation}
where $\mathcal{H}(X)=-\sum\nolimits_{x}p_{x}\log p_{x}$ is the usual Shannon
entropy for the variable $X$ and $\mathcal{H}(X,Y)=-\sum\nolimits_{x,y}%
p_{x,y}\log p_{x,y}$ is the corresponding joint entropy. Here $p_{x}$ is the
probability of variable $X$ to assume the value $x$ and $p_{x,y}\equiv
p(y|x)p(x)$, where $p(y|x)$ is the conditional probability of occurrence of
$y$ when $x$ has already occurred. Accordingly, Eq.~(\ref{CMI}) can be
straightforwardly extended for a bipartite quantum state described by the
density matrix $\rho_{AB}$ as follows~\cite{NieChu, Bennetti, Vedral-Book}:
\begin{equation}
\mathcal{I}(\rho_{A:B})=S(\rho_{A})+S(\rho_{B})-S(\rho_{AB}),\label{QMI}%
\end{equation}
where $S(\rho)=-\operatorname*{Tr}(\rho\log\rho)$ is the von Neumann entropy,
$\rho_{A}=\operatorname*{Tr}_{B}(\rho_{AB})$ is the reduced density operator
of the partition $A$, and accordingly for $\rho_{B}$. Equation~(\ref{QMI})
gives a measure of the total correlations (including the quantum and the
classical ones) contained in a bipartite quantum
system~\cite{Maziero,Maziero2,GroPoWi,SchuWest}. The classical part of
correlation~(\ref{QMI}) can be expressed as being the \textquotedblleft
maximum classical mutual information\textquotedblright\ obtained by local
measurements on both partitions of a composite state~\cite{DiVincenzo}:
\begin{equation}
\mathcal{K}(\rho_{AB})=\underset{\left\{  \Pi_{i}^{(A)}\otimes\Pi_{j}%
^{(B)}\right\}  }{\max}\left[  I_{c}(\rho_{A:B})\right]  .\label{CC}%
\end{equation}
Here $I_{c}(\rho_{A:B})$ is given by Eq.~(\ref{CMI}) provided that
$\mathcal{H}(X)$ is seen in this case as the entropy of the probability
distribution of system $X$ resulting from a set of local projective
measurements $\Pi_{i}^{(A)}\otimes\Pi_{j}^{(B)}$ on both subsystems $A$ and
$B$. The maximization is taken over the set of all possible projective measurements.

Due to the distinct nature of both classical and quantum correlations it is
reasonable to assume that they add in a simple
way~\cite{Maziero,Maziero2,GroPoWi,Horo1,Horo2}. Hence, we can define the
quantum part of the total correlation as
\begin{equation}
\mathcal{Q}(\rho_{AB})\equiv\mathcal{I}(\rho_{A:B})-\mathcal{K}(\rho
_{AB}).\label{QC}%
\end{equation}
Note that we are using here the \textquotedblleft two-side\textquotedblright%
\ correlation measure (in the sense that both subsystems are
measured)~\cite{Maziero2}, instead of the \textquotedblleft
one-side\textquotedblright\ correlation measure as assumed in the quantum
discord~\cite{OllZur}. This choice relies on the fact that the Alice-Rob
problem considered in this article is not symmetric under the permutation of
the subsystems, since Alice and Rob are inertial and noninertial, respectively.

For the sake of completeness, we compare the results obtained by means of
Eq.~(\ref{QC}) with those obtained through the quantum discord
\begin{equation}
\mathcal{D}(\rho_{AB})\equiv\mathcal{I}(\rho_{A:B})-\max_{\left\{  \Pi
_{j}^{(B)}\right\}  }\mathcal{J}(\rho_{A\text{:}B}).\label{Dis}%
\end{equation}
Here
\[
\mathcal{J}(\rho_{A:B})=S(\rho_{A})-S_{\left\{  \Pi_{j}^{(B)}\right\}
}\left(  \rho_{A|B}\right)  ,
\]
where $S_{\left\{  \Pi_{j}^{(B)}\right\}  }\left(  \rho_{A|B}\right)  $ is the
quantum extension of the classical conditional entropy $\mathcal{H}(A|B)$,
which (in the quantum case) depends on the measurement choice. Thus, to
compute the quantum discord, we have to maximize this quantity over the set
$\left\{  \Pi_{j}^{(B)}\right\}  $ of all possible projective measurements on
subsystem $B$. Of course, there is a related one-side measure of classical
correlation given by $\mathcal{C}(\rho_{AB})\equiv\mathcal{I}(\rho
_{A:B})-\mathcal{D}(\rho_{AB})$.%

\begin{figure}
[h]
\begin{center}
\includegraphics[
natheight=7.932100in,
natwidth=11.361000in,
height=2.2347in,
width=2.9914in
]%
{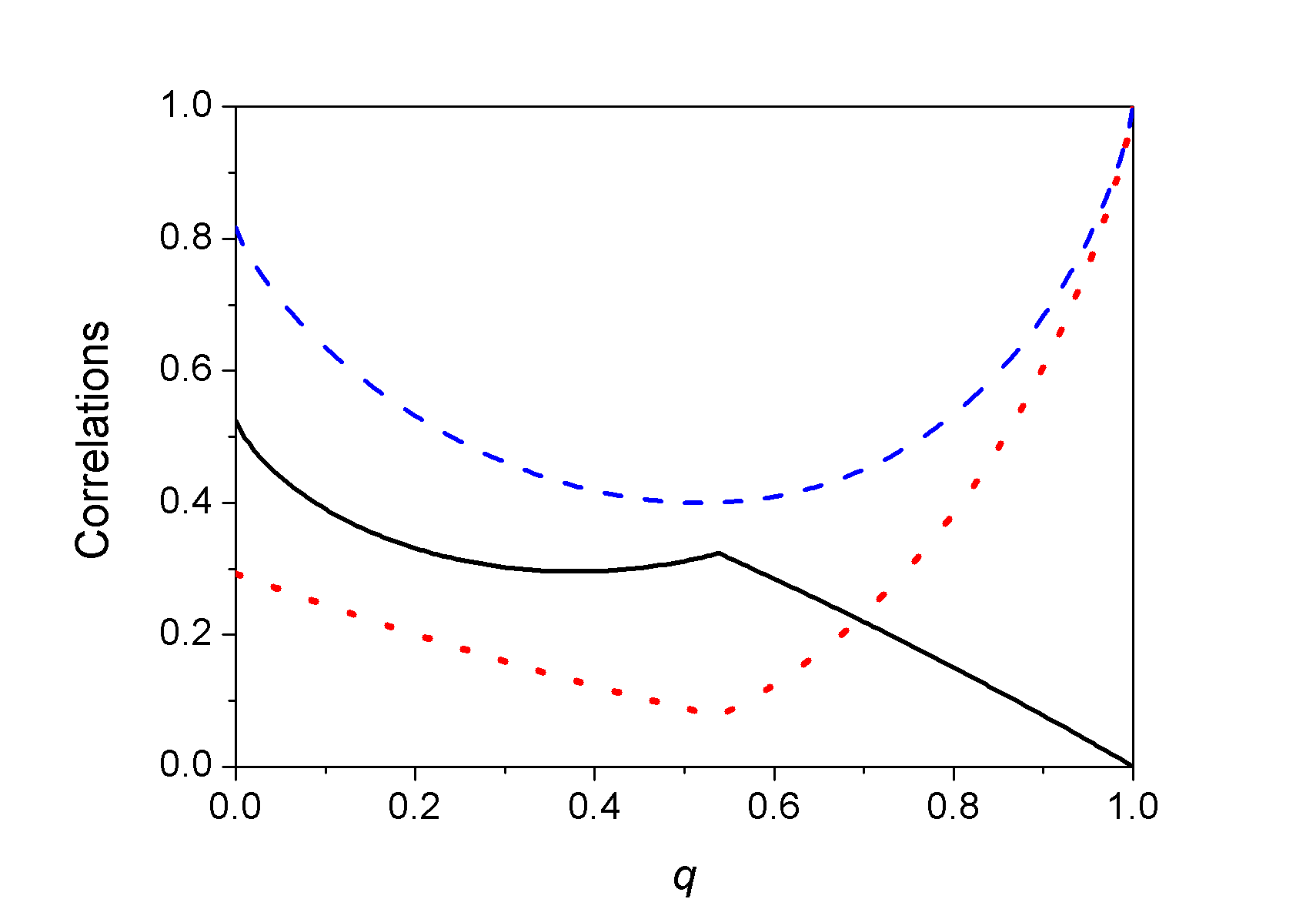}%
\caption{(Color online) We plot the classical (dotted line) and quantum (solid
line) correlations given by Eqs.~(\ref{CC}) and~(\ref{QC}), respectively, and
the quantum mutual information (dashed line) given by Eq.~(\ref{QMI}) as
functions of the parametrized acceleration $q$ considering the initial
state~(\ref{IS}). Here we set $\nu^{2}=0.4\,\pi$.}%
\label{Fig1}%
\end{center}
\end{figure}
%

\begin{figure}
[h]
\begin{center}
\includegraphics[
natheight=7.932100in,
natwidth=11.361000in,
height=2.2675in,
width=3.0381in
]%
{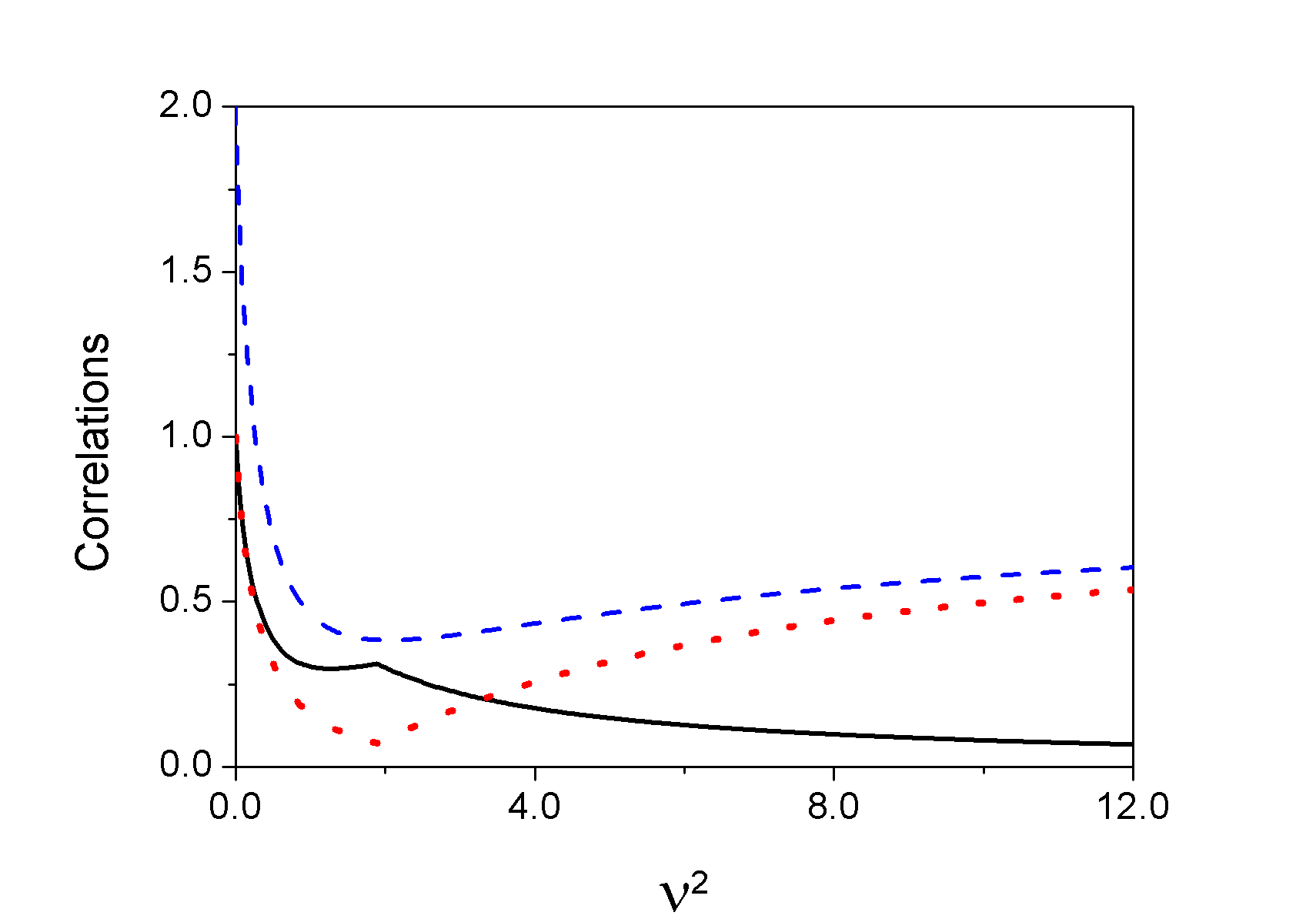}%
\caption{(Color online) We plot the same correlations as in Fig.~1 with
respect to $\nu^{2}$ assuming the initial state~(\ref{IS}) and $q=0.4$. The
quantum mutual information~(\ref{QMI}) is given by the dashed line and the
classical and quantum correlations~(\ref{CC}) and~(\ref{QC}) are given by the
dotted and solid lines, respectively.}%
\label{Fig2}%
\end{center}
\end{figure}


\section{Dynamics of correlations and the Unruh effect}

\label{Correlation dynamics}

\begin{figure}
[ptbh]
\begin{center}
\includegraphics[
natheight=4.179600in,
natwidth=5.000400in,
height=2.5452in,
width=3.039in
]%
{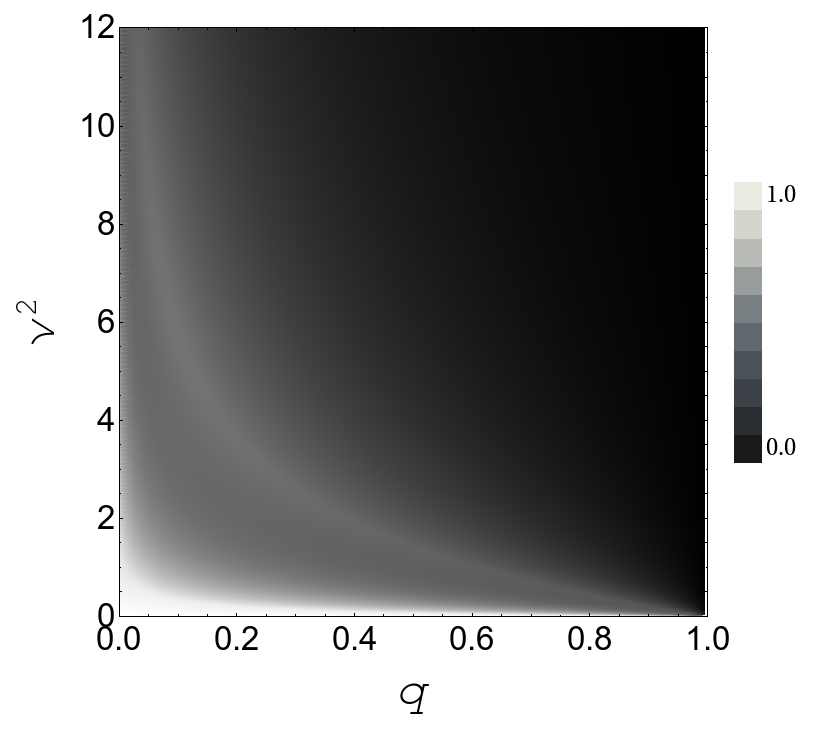}%
\caption{Density plot of quantum correlation given in Eq.~(\ref{QC}) as a
function of $\nu^{2}$ and $q$, considering the initial state~(\ref{IS}).}%
\label{Fig3}%
\end{center}
\end{figure}

\begin{figure}
[ptbh]
\begin{center}
\includegraphics[
natheight=4.479700in,
natwidth=5.360100in,
height=2.5443in,
width=3.039in
]%
{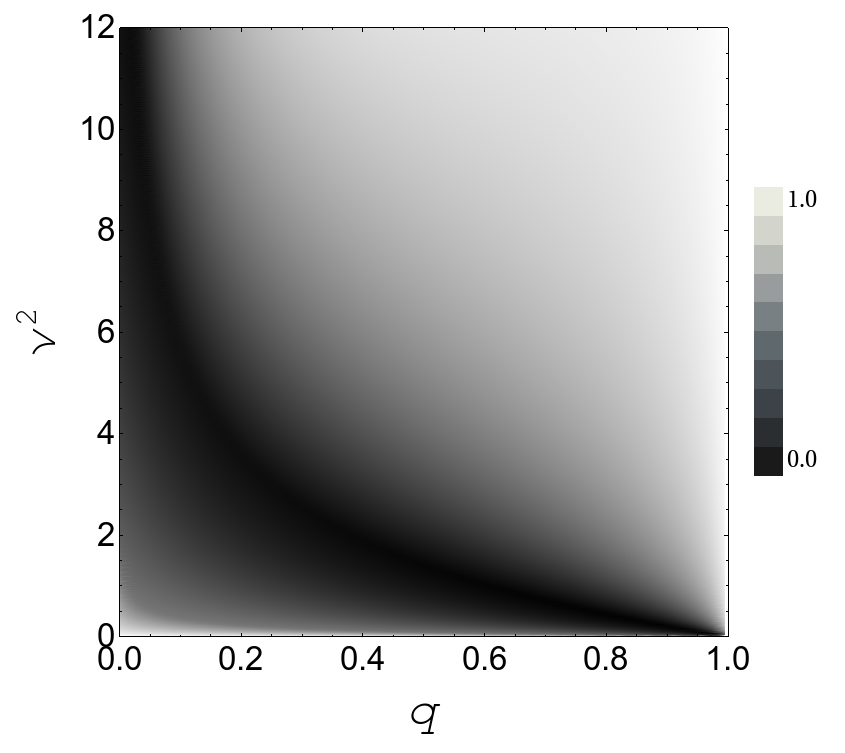}%
\caption{Density plot of classical correlation given in Eq.~(\ref{CC}) as a
function of $\nu^{2}$ and $q$, considering the initial state~(\ref{IS}).}%
\label{Fig4}%
\end{center}
\end{figure}

\begin{figure}
[ptbh]
\begin{center}
\includegraphics[
natheight=4.226300in,
natwidth=5.053100in,
height=2.5469in,
width=3.039in
]%
{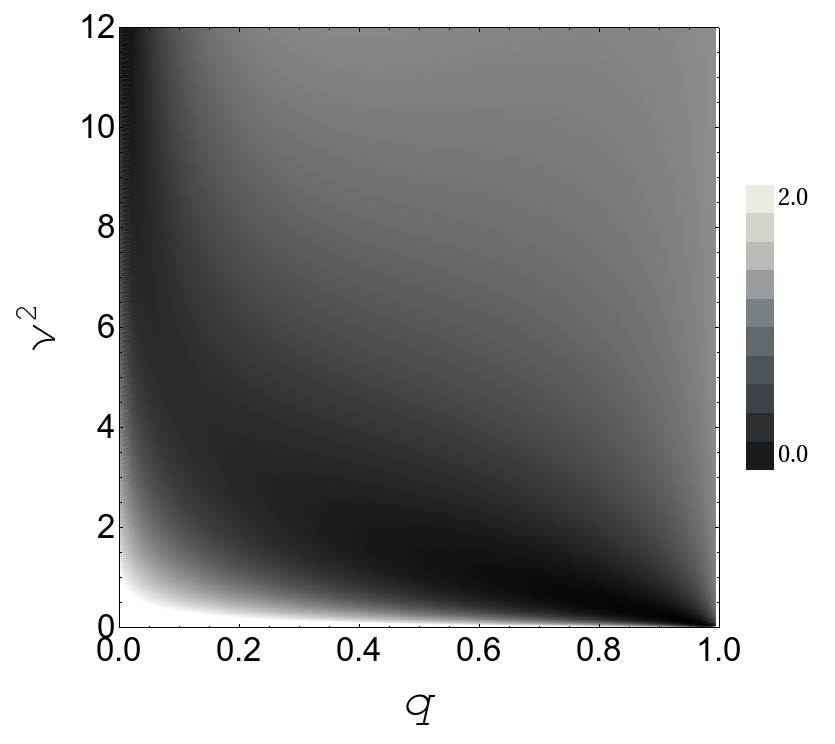}%
\caption{Density plot of the quantum mutual information given in
Eq.~(\ref{QMI}) as a function of $\nu^{2}$ and $q$, considering the initial
state~(\ref{IS}).}%
\label{Fig5}%
\end{center}
\end{figure}

Let us now proceed to analyze the behavior of both classical and quantum
correlations given by Eqs.~(\ref{CC}) and~(\ref{QC}), respectively, assuming
point detectors: $\kappa=0$. In Fig.~\ref{Fig1}, the quantum and classical
correlations as well as the mutual information are plotted as functions of the
parametrized acceleration $q$ for a fixed value of $\nu^{2}$ and time interval
$\Delta$ along which the detector stays switched on. First, we note that for
null acceleration the total correlation value given by the quantum mutual
information differs from $2$, as would be expected for the singlet
state~(\ref{IS}). This is so because even inertial detectors have a nonzero
probability of spontaneously decaying (along the nonzero time interval
$\Delta$) with the emission of a Minkowski particle, which carries away some
information. Such a process leads to a purity loss of the initial singlet
state degradating the initial correlations. It is worthwhile to emphasize that
the usual value of 2 for the quantum mutual information of the singlet state
is obtained when the effective coupling constant $\nu^{2}$ vanishes, as can be
seen from Fig. 2. In such a case there is no interaction between Rob%
\'{}%
s qubit and the scalar field. Figure~\ref{Fig1} shows that the two-qubit
system is still left with classical correlations in the limit of infinite
acceleration ($q\rightarrow1$), while the quantum correlation vanishes. This
fact is in agreement with Eq. (\ref{assimpt}) and, as mentioned before, this
is a consequence of our detector model.\textbf{ \ }From the point of view of
uniformly accelerated observers, the loss of quantum correlation is due to the
interaction of Rob's qubit with the Unruh thermal bath of Rindler particles
that they experience when the field is in the Minkowski vacuum. We recall that
the Unruh temperature experienced by the non-inertial qubit is proportional to
its proper acceleration. Now from the inertial observers' perspective, the
quantum correlation is carried away by the scalar radiation emitted by the
accelerating qubit when it suffers a transition. Another interesting result
revealed by Fig.~\ref{Fig1} is the fact that the dynamics of both classical
and quantum correlations cannot be described by a smooth function of
acceleration. Extensive numerical analyses indicate that this \emph{sudden
change} does not depend on the considered initial state. (The term
\textquotedblleft\emph{sudden change}" was coined in Ref.~\cite{Maziero} and
experimentally observed in Ref.~\cite{Guo}.) For larger values of $\nu^{2}$,
the sudden-change point moves to the left on the $q$ axis. A similar behavior
is observed when we plot the same correlations as functions of $\nu^{2}$ for a
fixed value of acceleration, as seen in Fig.~\ref{Fig2}. The larger the
acceleration, the closer to the origin the sudden-change point approaches. In
Fig.~\ref{Fig3}, we show how the sudden-change point depends on the parameters
$q$ and $\nu^{2}$. It becomes clear that the quantum correlation vanishes for
every $\nu^{2}$ value provided that the acceleration is arbitrarily large.
Figures~\ref{Fig4} and~\ref{Fig5} show the classical correlation~(\ref{CC})
and quantum mutual information~(\ref{QMI}), respectively. It is clear that for
arbitrarily large accelerations all correlations left in the two-qubit system
have a classical rather than a quantum nature for all $\nu^{2}$ values.%

%

%

\begin{figure}
[ptbh]
\begin{center}
\includegraphics[
natheight=7.932100in,
natwidth=11.361000in,
height=2.3359in,
width=2.9914in
]%
{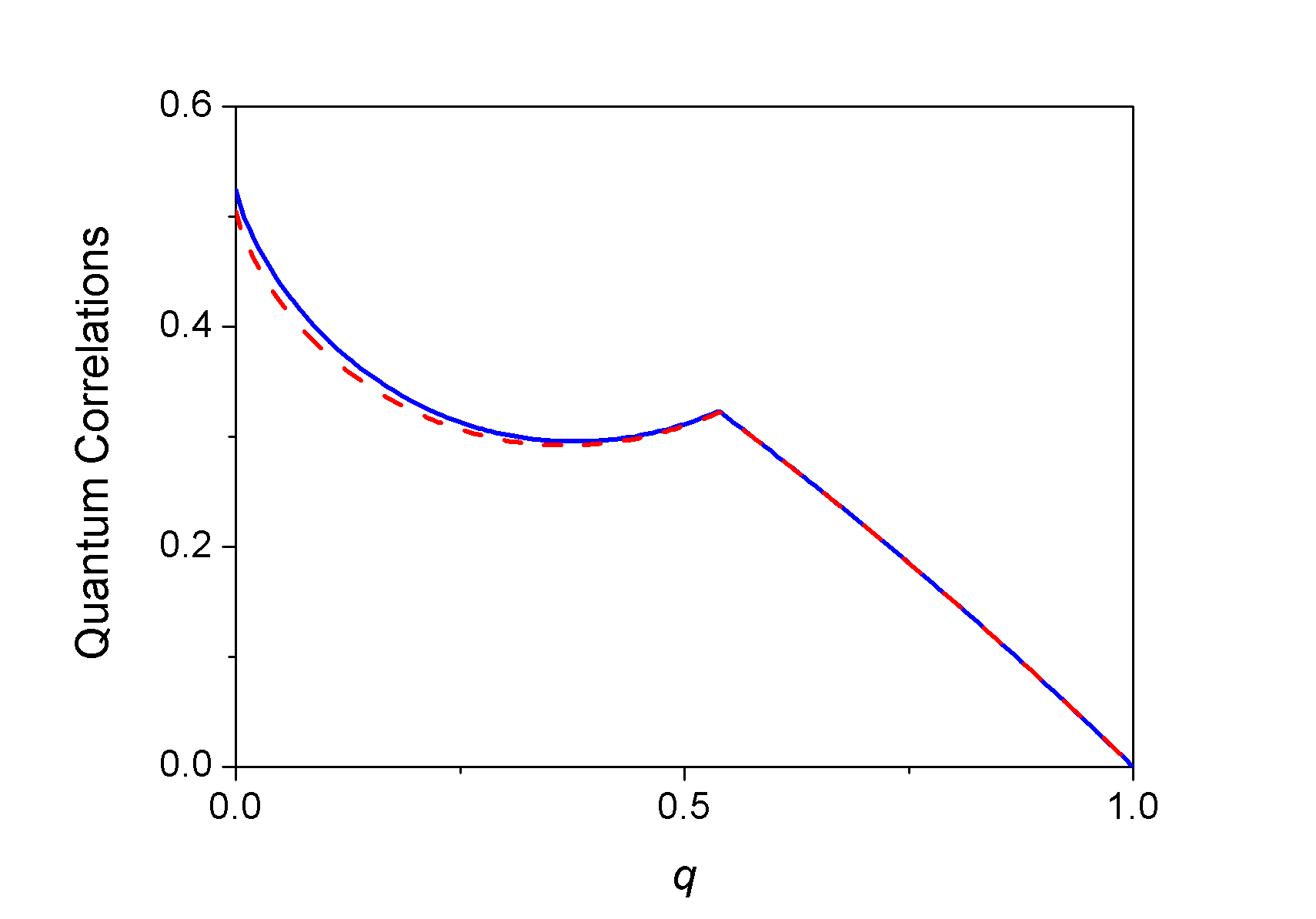}%
\caption{(Color online) Plot of quantum correlations as functions of the
parametrized acceleration $q$, setting $\nu^{2}=0.4\,\pi$. The symmetrical
measure~(\ref{QC}) is shown by the solid line, while the quantum
discord~(\ref{Dis}) is given by the dashed line (for measurements performed by
either Alice or Rob).}%
\label{Fig6}%
\end{center}
\end{figure}

As stated in Sec.~\ref{Introduction}, there is no \textit{a priori} reason to
believe that the two quantifiers of quantum correlation, given by
Eqs.~(\ref{QC}) and~(\ref{Dis}), should lead to the same results for a system
that is not symmetrical under permutation of its parts. Indeed, in our case
they are shown to be distinct although close to each other (see
Fig.~\ref{Fig6}).
\begin{figure}
[ptbh]
\begin{center}
\includegraphics[
natheight=8.684400in,
natwidth=11.403400in,
height=2.2909in,
width=2.9386in
]%
{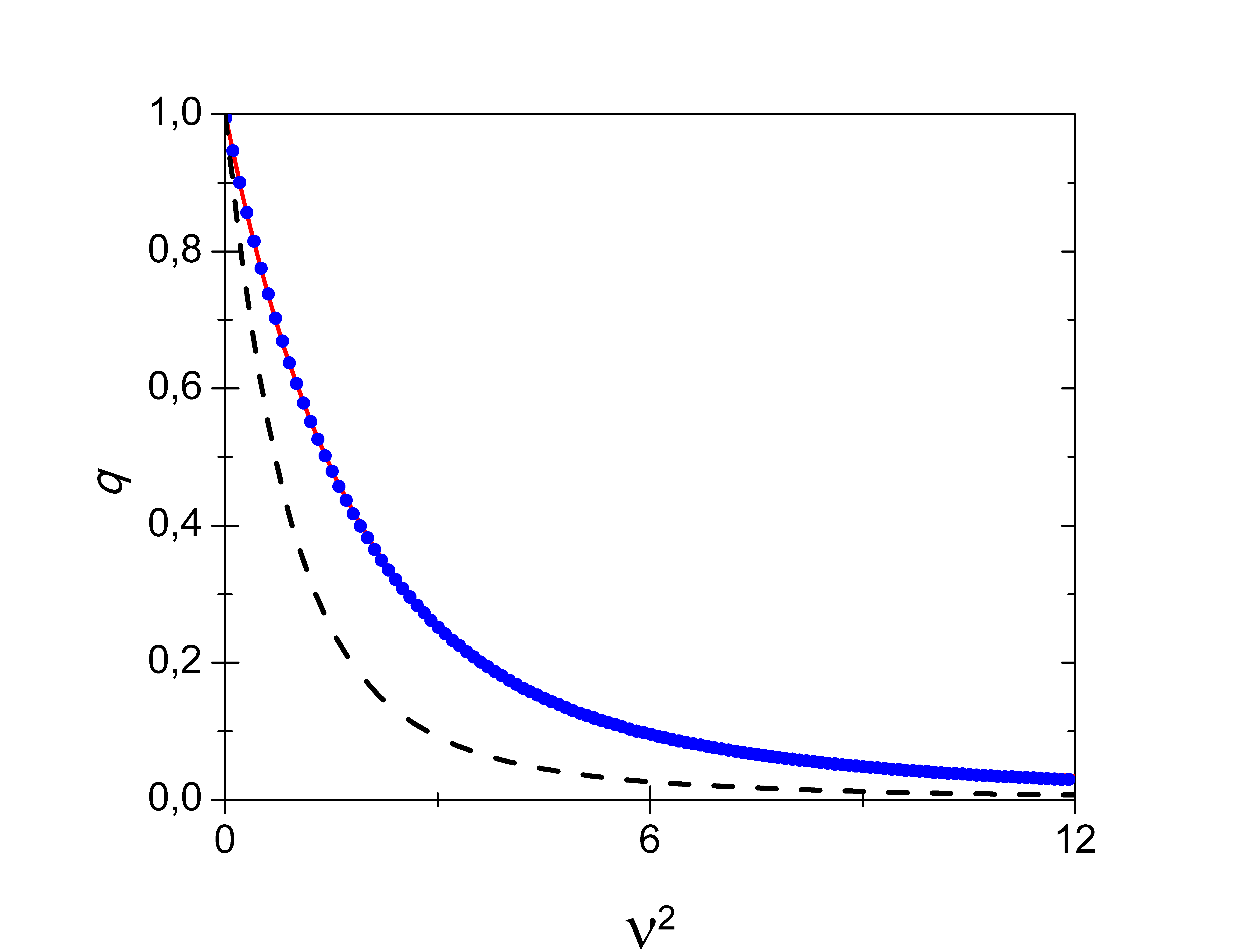}%
\caption{(Color online) The dotted line shows the plot of the numerical
solutions of Eq.~(\ref{TE}). The solid line is the exponential fitting of
these solutions. The dashed line is the plot of Eq.~(\ref{ESD}), showing the
values of $q$ and $\nu^{2}$ for which the entanglement sudden death occurs.}%
\label{Fig7}%
\end{center}
\end{figure}
An interesting fact that can be also seen from Fig.~\ref{Fig6} is that both
quantifiers appear to give the same value for the \emph{sudden-change} point
$q_{sc}$. This allows us to derive in the present case an approximate
analytical expression for $q_{sc}$ as follows. In a recent
article~\cite{Fanchini}, it was shown that for all bipartite density operators
of the form
\begin{equation}
\rho_{AB}=\left[
\begin{array}
[c]{cccc}%
\rho_{11} & 0 & 0 & \rho_{14}\\
0 & \rho_{22} & \rho_{23} & 0\\
0 & \rho_{23} & \rho_{22} & 0\\
\rho_{14} & 0 & 0 & \rho_{33}%
\end{array}
\right]  \label{Form}%
\end{equation}
the quantum discord~(\ref{Dis}) is given by
\begin{equation}
\mathcal{D}=\min\{D_{1},D_{2}\}\;\;\;(D_{1},D_{2}\geq0),\label{Disa}%
\end{equation}
where
\begin{align}
D_{1} &  =S(\rho_{A})-S(\rho_{AB})-\rho_{11}\log\left[  \frac{\rho_{11}}%
{\rho_{11}+\rho_{22}}\right]  \nonumber\\
&  -\rho_{22}\log\left[  \frac{\rho_{22}^{2}}{\left(  \rho_{11}+\rho
_{22}\right)  \left(  \rho_{33}+\rho_{22}\right)  }\right]  \nonumber\\
&  -\rho_{33}\log\left[  \frac{\rho_{33}}{\rho_{33}+\rho_{22}}\right]  ,
\end{align}%
\begin{align}
D_{2} &  =S(\rho_{A})-S(\rho_{AB})-\frac{1}{2}\left(  1+\Gamma\right)
\log\left[  \frac{1}{2}\left(  1+\Gamma\right)  \right]  \nonumber\\
&  -\frac{1}{2}\left(  1-\Gamma\right)  \log\left[  \frac{1}{2}\left(
1-\Gamma\right)  \right]  ,
\end{align}
and $\Gamma^{2}\equiv(\rho_{11}-\rho_{33})^{2}+4|\rho_{23}+\rho_{14}|^{2}$. By
using the density operator $\rho_{\infty}^{AR}$ of Eq.~(\ref{FS}) in
Eq.~(\ref{Form}), we obtain
\begin{align*}
D_{1} &  =2S_{0},\\
D_{2} &  =S(\rho_{A})-S(\rho_{AB})-\frac{1}{2}\log\left[  \frac{1}{4}%
(1-\Gamma^{2})\right]  \\
&  +\frac{1}{2}\Gamma\log\left[  \frac{1-\Gamma}{1+\Gamma}\right]
\end{align*}
with $\Gamma^{2}=(S_{2}-S_{1})^{2}+4S_{0}^{2}$. Then, we see from
Eq.~(\ref{Disa}) that the \emph{sudden-change} point is obtained by solving
\begin{equation}
D_{1}=D_{2},\label{TE}%
\end{equation}
which turns out to be a transcendental equation for $q$ and $\nu^{2}$. For the
case plotted in Fig.~\ref{Fig1}, Eq.~(\ref{TE}) is satisfied for
$q=q_{sc}\approx0.53925$, which is in agreement with the numerical analysis
presented in the last section. An approximate semi-analytical expression for
$q_{sc}$ can be found as follows. First, we find numerically all pairs
($q,\nu^{2}$) which satisfy~Eq. (\ref{TE}). Next, an exponential fitting
\begin{equation}
q_{sc}=\exp[a+b\nu^{2}+c\nu^{4}]\label{EF}%
\end{equation}
is performed on these solutions with parameters
\[
a=0.00054,\;\;\;b=-0.51488,\;\;\;c=0.01959.
\]
As we can see from Fig.~\ref{Fig7}, Eq.~(\ref{EF}) is in very good agreement
with the solutions of Eq.~(\ref{TE}) for the range of parameters considered
here. Figure~\ref{Fig7} also shows the acceleration values $q_{sd}$ for which
the entanglement sudden death~\cite{Yu} occurs as given in Ref.~\cite{LM209}:
\begin{equation}
q_{sd}=(\nu^{2}/2+\sqrt{1+\nu^{4}/4})^{-2},\label{ESD}%
\end{equation}
which suggests that the entanglement sudden death and the quantum correlation
\emph{sudden change} are uncorrelated. Furthermore, we note that the quantum
correlation as defined by Eqs.~(\ref{QC}) and~(\ref{Dis}) is still present
after the vanishing of the entanglement.%

\begin{figure}
[ptb]
\begin{center}
\includegraphics[
natheight=8.700000in,
natwidth=11.375700in,
height=2.4059in,
width=2.936in
]%
{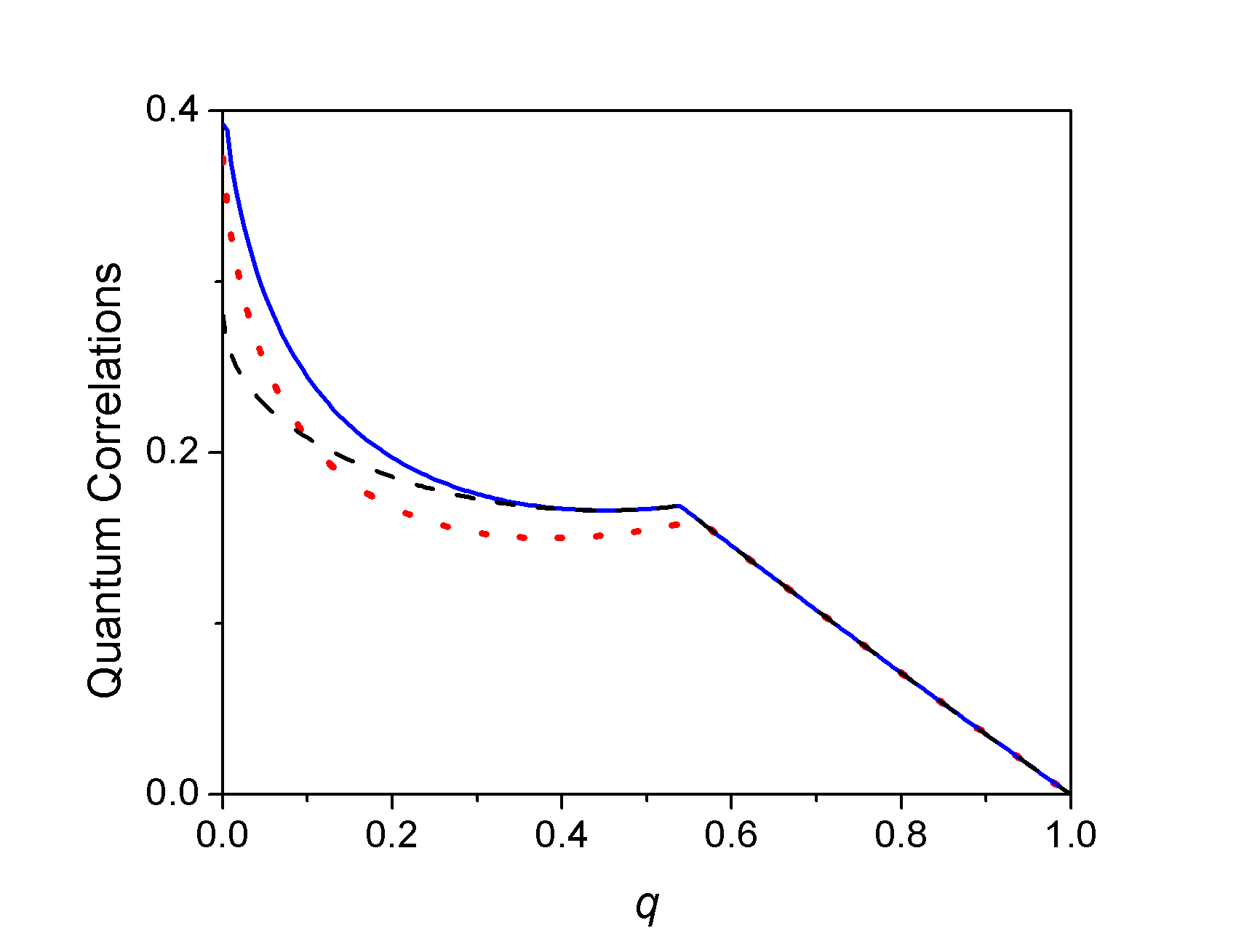}%
\caption{(Color online) Plot of quantum correlations as a function of the
parametrized acceleration $q$, setting $\nu^{2}=0.4\,\pi$. The symmetrical
measure~(\ref{QC}) is shown by the solid line, while the quantum
discord~(\ref{Dis}) is given by the dotted line (for a measurement performed
by Rob) and by the dashed line (for a measurement performed by Alice).}%
\label{Fig8}%
\end{center}
\end{figure}

\begin{figure}
[ptbh]
\begin{center}
\includegraphics[
natheight=5.199200in,
natwidth=6.679800in,
height=2.7138in,
width=3.4791in
]%
{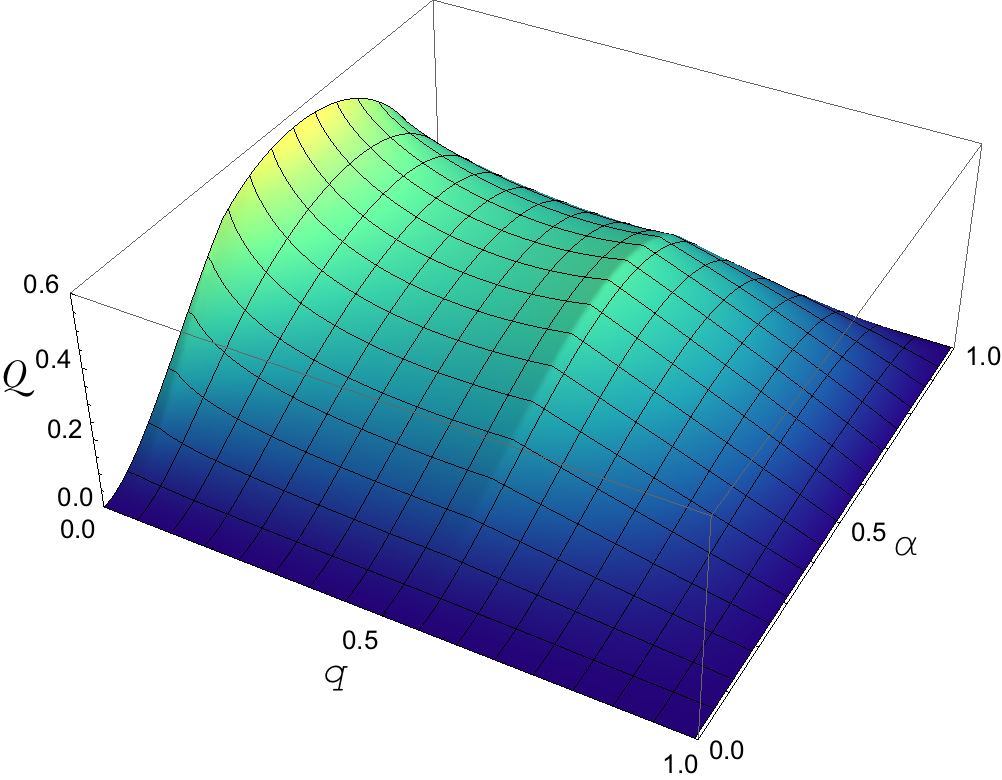}%
\caption{(Color online) The quantum correlation given in Eq.~(\ref{QC}) is
plotted as a function of $\alpha$ and $q$ setting $\nu^{2}=0.4\,\pi$ for the
initial state~(\ref{IS2}). We note that it exhibits a sudden change across the
$q\approx0.54$ line.}%
\label{Fig9}%
\end{center}
\end{figure}

Now, let us wonder what happens if we consider a nonsymmetrical initial state
\begin{equation}
|\Psi_{0}^{\prime\,AR\phi}\rangle=(\alpha|0_{A}\rangle\otimes|1_{R}%
\rangle-\beta|1_{A}\rangle\otimes|0_{R}\rangle)\otimes|0_{M}\rangle\label{IS2}%
\end{equation}
under permutation of the $A$ and $R$ qubits, where $|\alpha|^{2}+|\beta
|^{2}=1$ rather than the symmetrical one~(\ref{IS}). In Fig.~\ref{Fig8}, we
plot the quantum discord~(\ref{Dis}) and the two-side quantum
correlation~(\ref{QC}) assuming the state~(\ref{IS2}) with $\alpha=0.3$. As we
can see, the quantum discord for the two partners, Alice (inertial) and Rob
(noninertial), differs from the two-side measure [Eq. (\ref{QC})] and from
each other, leading to distinct \emph{sudden-change} points for the two
observers. In such situations, using symmetrical correlation measures seems to
be more suitable. For the cases considered in this article, the \emph{sudden
changes} associated with (i)~the quantum discord as computed by the inertial
observer and (ii)~the two-side measurement always seem to agree with each
other. However, Eq.~(\ref{TE}) is not valid anymore due to the fact that the
density operator for the Alice-Rob system is not in the form given by
Eq.~(\ref{Form}). Figure~\ref{Fig9} shows a plot of Eq.~(\ref{QC}) for the
initial state~(\ref{IS2}), where we can see that the \emph{sudden-change}
point does not seem to vary with $\alpha$ (and $\beta$). Moreover, this
critical point always exists except for the separable pure states, where
$\alpha=0$ or~$1$.
The (omitted) plot for the quantum discord as measured by the inertial
observer $A$ leads to a similar behavior with agreement on the
\emph{sudden-change} point.


\section{Classical and Quantum Correlations in the vicinity of a Black Hole}

\label{Black hole}

\begin{figure}
[ptb]
\begin{center}
\includegraphics[
natheight=3.968600in,
natwidth=6.499900in,
height=1.8836in,
width=3.077in
]%
{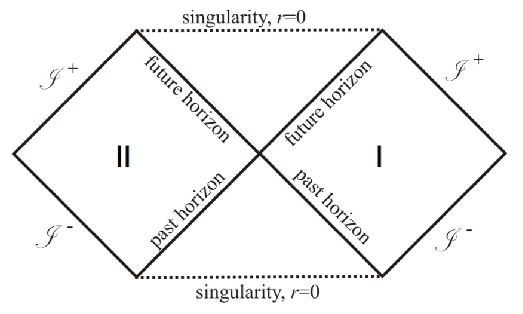}%
\caption{Extended Schwarzschild space-time.}%
\label{Fig10}%
\end{center}
\end{figure}

Let us now explain how our previous results translate when the two qubits are
very close to a Schwarzschild (i.e., rotationless and chargeless) black hole.
First, let us recall that in this case there is a static Killing field $\chi$
which is timelike in regions $I$ and $II$ of the extended Schwarzschild
space-time (see Fig.~\ref{Fig10}). The Hartle-Hawking vacuum is the unique
nonsingular (Hadamard) state, which is invariant under the isometries
generated by $\chi$~\cite{wald94, KW91, HH76, I76, K85}.
By a similar analysis used to derive the Unruh effect, it can be shown that
the Hartle-Hawking vacuum $|0_{HH}\rangle$ is a thermal state with respect to
the time translation generated by $\chi$ when restricted to region $I$. This
should be identified with the so-called Hawking radiation as seen by static
observers lying outside the black hole. The corresponding reduced density
matrix can be cast as~\cite{wald94, I76}
\begin{equation}
\rho_{HH}=\prod_{i}(C_{i}\sum_{n_{iI}}e^{-2\pi n_{iI}\omega_{i}/\kappa}%
|n_{iI}\rangle\langle n_{iI}|),
\end{equation}
where $\kappa=1/4M$ is the surface gravity, $M$ is the black hole mass,
$C_{i}=(1-e^{-2\pi\omega_{i}/\kappa})^{\frac{1}{2}}$, and $|n_{iI}\rangle$ are
states with $n$-particles in the modes $f_{i}^{I}$. Here $f_{i}^{I}$ and
$f_{i}^{I\ast}$ consist of a complete set of orthonormal $\chi$
positive-frequency solutions of the Klein-Gordon equation in region $I$ and
vanish in region $II$. We emphasize that, as a consequence both (i) uniformly
accelerated observers in Minkowski vacuum $|0_{M}\rangle$ and (ii) static
observers outside the black hole (who also have constant proper acceleration)
in the Hartle-Hawking vacuum $|0_{HH}\rangle$ experience a thermal bath of all
particles. In contrast, free-falling observers in Minkowski space-time and in
the vicinity of the black hole with the corresponding $|0_{M}\rangle$ and
$|0_{HH}\rangle$ vacua, respectively, see no particles at all~\cite{Unruh}.

Now, we determine the precise setup for Alice and Rob which allows the
translation of our previous results for the black hole case. For this purpose,
let us consider for the sake of simplicity the line element of a
two-dimensional Schwarzschild black hole:
\begin{equation}
ds^{2}=-(1-2M/r)dt^{2}+(1-2M/r)^{-1}dr^{2},
\end{equation}
where $(t,r)$ are the usual static Schwarzschild coordinates. By making the
change of variables $r\rightarrow\rho(r)=\sqrt{8M(r-2M)}$, we see that very
close to the horizon, $r\approx2M,$ the metric takes the form
\begin{equation}
ds^{2}=-(\rho/4M)^{2}dt^{2}+d\rho^{2}.\label{rindler}%
\end{equation}
This is the line element of the Rindler wedge provided that $0\leq\rho<\infty$
and $-\infty<t<\infty$. The Rindler wedge is the section of the Minkowski
space-time which is covered by uniformly accelerated observers, who see the
Unruh thermal bath when the field state is in the Minkowski vacuum. Then, very
near the event horizon, the Schwarzschild and Rindler space-times resemble
each other. The fact that they differ asymptotically is not important provided
that the qubits are localized near the horizon. As a matter of fact, this
conclusion holds exactly the same for four-dimensional physical black holes.
The \textquotedblleft large" proper accelerations $a$ experienced by static
observers near the black hole horizon are associated with small time scales in
comparison with $r_{H}=2M$, making any curvature effects negligible for our
purposes. The local temperature measured by the static observers is given by
$T=\kappa/2\pi V$ where $V=(-g(\chi,\chi))^{1/2}$ is the redshift factor and
the surface gravity can be cast as $\kappa=\lim_{\mathrm{horizon}}(Va)$. Then,
the temperature experienced by static observers very close to the horizon in
the Hartle-Hawking vacuum is $T=a/2\pi$. This is analogous to the Unruh
temperature for observers with constant proper acceleration $a$ in Minkowski vacuum.

As a result, all the conclusions of the previous sections continue to be valid
in the vicinity of a Schwarzschild black hole with Hawking radiation provided
that Alice is free falling and Rob is static with the same proper acceleration
$a$ as if it were in Minkowski space-time. In particular, entanglement sudden
death would be seen at finite acceleration $a_{sd}$~\cite{LM209}, while the
quantum correlation would be completely destroyed only for $a\rightarrow
\infty$ (i.e., when Rob is arbitrarily close to the black hole horizon). The
\emph{sudden-change} phenomenon would be observed accordingly at the same
$q=q_{sc}$ values.


\section{Final Remarks}

\label{Final}

Here, we have analyzed the behavior of quantum and classical correlations for
a two-qubit system where one of them is uniformly accelerated while the other
one is inertial. We have shown that the quantum correlation is degraded by the
presence of acceleration. From the point of view of inertial observers,
uniformly accelerated qubits have a nonzero probability of exciting with the
emission of a Minkowski particle. This is possible because the external
accelerating agent provides the necessary energy. Now, owing to the Unruh
effect, coaccelerated observers with the qubit experience the Minkowski vacuum
as a thermal bath (of Rindler particles). From their point of view, the qubit
is in contact with the Unruh thermal bath and the excitation is because of the
absorption of a particle from this hot reservoir. Clearly the physical
observables encoded in the correlations do not depend on the particular
observer description. In the limit of infinite acceleration, the Unruh thermal
bath has an arbitrarily large temperature and it is natural to expect that
quantum correlations be completely destroyed; however, classical correlations
are still left \textbf{[}see Eq.~(\ref{assimpt})\textbf{]}. Another
interesting point is that the behavior of both classical and quantum
correlations cannot be described by smooth functions of the acceleration. They
present a \emph{sudden change} for a critical value of $q=q_{sc}$. By
comparing the quantum discord with a symmetrical quantifier for the quantum
correlation, Eq. (\ref{QC}), we found that although both measures are not
exactly the same, they are very close to each other, when we assume a
symmetrical initial singlet state. This fact has allowed us to find an
analytical fitting for $q_{sc}$ in terms of $\nu^{2}$. Our results indicate
that the correlation \emph{sudden-change} and entanglement sudden-death points
are uncorrelated in this scenario. Finally, by considering a nonsymmetrical
rather than symmetrical initial state, we found that the quantum discord and
the two-side measure for the quantum correlation become quite distinct. Also,
the quantum discord as computed by the inertial $A$ and noninertial $R$
observers are distinct leading to different \emph{sudden-change} points
$q_{sc}$. This result suggests that symmetrical measures should be more
suitable to be used to evaluate the correlations among the parties when the
experimentalists have distinct inertial features as it would be the case when,
for example, one of them is free falling and the other one is static in a
gravitational field. Eventually the precise prescription of how our results
can be translated when Alice is free falling and Rob is static in the vicinity
of a Schwarzschild black hole is given.

\begin{acknowledgments}
L.C. and A.L. acknowledge full financial support from Funda\c{c}\~{a}o
Universidade Federal do ABC and Funda\c{c}\~{a}o de Amparo \`{a} Pesquisa do
Estado de S\~{a}o Paulo (FAPESP), respectively. R.S. is grateful to Conselho
Nacional de Desenvolvimento Cient\'{\i}fico e Tecnol\'{o}gico (CNPq),
Instituto Nacional de Ci\^{e}ncia e Tecnologia em Informa\c{c}\~{a}o
Qu\^{a}ntica, and FAPESP for partial support. G.M. was partially supported by
CNPq and FAPESP.
\end{acknowledgments}

\end{document}